\documentclass[12pt]{article}

\catcode`\@=11
\@addtoreset{equation}{section}

\global\arraycolsep=2pt
\usepackage{amsbsy,amssymb,latexsym,amsfonts,amsmath}
\usepackage{graphicx,color}

\allowdisplaybreaks

\begin{document}
\begin{flushright}
\parbox{4.2cm}
{UCB-PTH-09/20}
\end{flushright}

\vspace*{0.7cm}

\begin{center}
{\Large \bf 
Gravity Dual for Reggeon Field Theory 

and Non-linear Quantum Finance
}
\vspace*{2.0cm}\\
{Yu Nakayama}
\end{center}
\vspace*{-0.2cm}
\begin{center}
{\it Berkeley Center for Theoretical Physics, \\ 
University of California, Berkeley, CA 94720, USA
}
\vspace{3.8cm}
\end{center}

\begin{abstract} 
We study scale invariant but not necessarily conformal invariant deformations of non-relativistic conformal field theories from the dual gravity viewpoint. We present the corresponding metric that solves the Einstein equation coupled with a massive vector field. We find that, within the class of metric we study, when we assume the Galilean invariance, the scale invariant deformation always preserves the non-relativistic conformal invariance.
We discuss applications to scaling regime of Reggeon field theory and non-linear quantum finance. These theories possess scale invariance but may or may not break the conformal invariance, depending on the underlying symmetry assumptions.
\end{abstract}

\thispagestyle{empty} 

\setcounter{page}{0}

\newpage

\section{Introduction} 
The invasion of gauge/gravity correspondence into other areas of theoretical physics is quite rapid these days. We have proliferating numbers of gauge/gravity correspondence beyond the classical AdS/CFT correspondence, first in AdS/QCD (like quark-gluon plasma applications) and then in AdS/CMP (like cold atom systems and high $T_c$ superconductors). In these examples, the gravity dual approach has given birth to completely novel models of the conventional physics. In particular, relatively new advancement in the gauge/gravity correspondence is to jump off from the island of the relativistic dispersion relation and dive into the vast ocean of non-relativistic systems \cite{Son:2008ye}\cite{Balasubramanian:2008dm}.
In this paper, we attempt to expand the realm of the gauge/gravity correspondence further by studying yet another application to  non-relativistic systems--- Reggeon field theory and quantum finance.

In the free field theory limit, the both systems are reduced to the free Schr\"odinger equation, so they have the right symmetry that could fit in the non-relativistic conformal invariant theories to begin with, albeit they are free. Once the interaction has been introduced, they are driven into the strongly coupled regime in the long distance (and over the long time). Nobody has achieved non-perturbative understanding of their strongly coupled regime, and our hope is that the gravitational dual approach may shed a new light on the dynamics.

Technically speaking, we have a completely orthogonal motivation for this study. In the relativistic conformal field theories, there is a long standing  question whether the scale invariance suggests conformal invariance or not. Group theoretically, there is no reason why the scale invariance indicates the conformal invariance because the former is merely a subgroup of the latter. Surprisingly, however, in the relativistic conformal field theories living in higher than two space-time dimension, there are no known examples of unitary scale invariant field theories but not conformal invariant. In two space-time dimension, under certain mild assumptions such as unitarity and discreteness of spectrum, it has been proved that the scale invariance indeed implies the conformal invariance \cite{Polchinski:1987dy}. When we remove a certain assumption such as unitarity, there are known examples of scale invariant but non-conformal field theories. See \cite{Ho:2008nr} for a recent construction of a non-trivially interacting example, which may be important in time-dependent string theories.  

It is of interest to ask the same question in the non-relativistic case. It is easy to construct scale invariant but non-conformal field theories, as we will encounter in this paper, once the Galilean invariance is broken. A part of the argument simply comes from the group structure because the non-relativistic conformal invariance demands $i[K,P_i] = G_i$, where $K$ is the non-relativistic conformal transformation, so once the Galilean boost $G_i$ is broken, one cannot preserve $K$ as long as momentum $P_i$ is conserved (see appendix A and B for more discussions). The remaining task is to find a scale invariant but Galilean non-invariant theory. An example is the Lifshitz-like field theory whose gravity counterpart has been studied in \cite{Kachru:2008yh}. We will see more examples from the gravity viewpoint in the main part of the paper.

The central question is, therefore, whether Galilean and scale invariant field theories are automatically non-relativistic conformal invariant or not. We cannot give a full account of this statement nor will we give any counterexamples in this paper. We will see, however, within the dual gravity background, the
deformations of the non-relativistic conformal field theory that preserve the Galilean invariance as well as scale invariance do not break the conformal invariance, either.

For applications to Reggeon field theory and non-linear quantum finance in the strongly coupled scaling regime, we notice that they possess the scale invariance but may or may not break the conformal invariance, depending on the underlying symmetry assumptions. Accordingly, we choose the gravity background in order to reproduce qualitative features (e.g. symmetries) of the theories we would like to study. In addition, we see that the Reggeon field theory with a simple cubic interaction shows a non-trivial renormalization group flow of the dynamical critical exponent. The tree level dynamical critical exponent $\mathcal{Z}=2$ will be modified through the interaction and the subsequent renormalization. The renormalization of the dynamical critical exponent (also known as the Hurst exponent) might be of relevance in quantum finance, too.

The organization of the paper is as follows. In section 2, we study various scale invariant but non-conformal deformations of the gravity solutions studied in literatures. We show that once Galilean invariance is imposed, the conformal invariance cannot be broken within the class of models we consider. We also compute the correlation functions in the deformed background. In section 3, we discuss the applications to Reggeon field theory and non-linear quantum finance. We have included a short review of the supersymmetric quantum mechanical formulation of the Black-Scholes-Merton model in section 3.2 to motivate the non-linear extension in section 3.3. We give further discussions and speculations in section 4. In appendix A, we study the criterion when the scale invariance and Galilean invariance indicates the conformal invariance from the field theory viewpoint. Appendix B summarizes the non-relativistic conformal algebra.
In appendix C, we collect some formulae for confluent Hypergeometric function used in the main text.

\section{Gravity Solutions}
In this section, we investigate scale invariant deformations of the gravity background proposed in \cite{Son:2008ye}\cite{Balasubramanian:2008dm}\cite{Kachru:2008yh}. A particular class of the metric satisfies the Einstein equation with massive vector field as a source of energy-momentum tensor. We study the correlation functions of the dual field theories by using the AdS/CFT technique.

\subsection{Scale invariant deformation of the non-relativistic conformal background}
Our starting point is the phenomenological gravity background for non-relativistic conformal field theories presented in \cite{Son:2008ye}\cite{Balasubramanian:2008dm}:
\begin{align}
ds^2_{d+2} = -2\frac{dt^2}{z^4} +\frac{-2dtd\zeta + dx_i^2 +dz^2}{z^2} \ , \label{nr}
\end{align}
where $i = 1, \cdots, d$. In later applications, we set $d=2$ for (physical) Reggeon field theory and $d=1$ for (one-factor) quantum finance. Originally, it was proposed to describe the cold atoms at criticality, or unitary fermion system for $d=3$ \cite{Son:2008ye}. 

The metric \eqref{nr} has the following isometries
\begin{itemize}
	\item Translations in $x_i$ and $t$.
	\item Rotations of $x_i$.
	\item Galilean boost: 

$(\zeta,x_i) \to \left(\zeta-v_i x_i+\frac{1}{2}v^2 t,x_i - v_it\right)$ .

	\item particle number (translation in $\zeta$).
	\item Scale transformation with dynamical critical exponent $\mathcal{Z} = 2$: 

$(t,\zeta,x_i,z) \to (\lambda^2t, \zeta, \lambda x_i, \lambda z)$.
	\item Non-relativistic special conformal transformation: 

$(t,\zeta,x_i,z) \to \left(\frac{t}{1+\eta t},\zeta-\frac{\eta}{2}\frac{x_ix_i+z^2}{1+\eta t}, \frac{x_i}{1+\eta t},\frac{z}{1+\eta t} \right)$ .
\end{itemize}
We will compactify $\zeta$ as $\zeta \simeq \zeta + 2\pi R$.

We would like to study scale invariant but not necessarily conformal invariant deformations of the metric \eqref{nr}. By assuming the scale invariance and translation invariance, possible metric deformations are restricted in the following three classes:
\begin{itemize}
	\item Class 1: The deformations that break the spatial rotation: $\delta ds^2 =  \left\{ \frac{dx_i dt}{z^3}, \frac{dx_i dz}{z^2}, \frac{dx_i d\zeta}{z} \right\}$. It also breaks the spatial parity invariance $x_i \to - x_i$. This kind of spatial anisotropy might be relevant in cosmology or condensed matter physics. Our applications in section 3, however, will preserve this symmetry, so we will not discuss the consequence of these deformations further in this paper.\footnote{Notice that the deformation $\frac{a_idx_i dz}{z^2}$ actually does not break the spatial rotation, so strictly speaking it is in class 3. By defining a new coordinate $x'_i = x_i - a_i z$, we can see the spatial rotation is realized in a disguised way.}
	\item Class 2: The deformations that break the Galilean boost: $\delta ds^2 = \left\{d\zeta^2\right\}$. The deformation also breaks the non-relativistic conformal invariance. The resulting continuous symmetry is same as that of the Lifshitz-like field theory (except for the presence of additional $U(1)$ particle number conservation). 

	\item Class 3: The deformations that preserve the Galilean boost: $\delta ds^2 = \left\{\frac{dzd\zeta}{z}, \frac{dzdt}{z^3}\right\}$. Actually, the deformation preserve the full non-relativistic conformal invariance as we will see shortly.
\end{itemize}

We first begin with how Class 3 deformations preserve the full non-relativistic conformal invariance. The key point is that the deformed metric is locally diffeomorphic to the original non-relativistic conformal invariant metric \eqref{nr}. To see this, we introduce the coordinate transformation
\begin{align}
t \to t + \frac{\alpha}{2}z^2 \ , \ \ \zeta \to \zeta + \beta \log z \ 
\end{align}
to the metric \eqref{nr}. The resulting metric in the new coordinate is
\begin{align}
ds^2_{d+2} = -2\frac{dt^2}{z^4} -(4\alpha + 2\beta) \frac{dt dz}{z^3} +\frac{-2dtd\zeta + dx_i^2 +(1-2\alpha^2-2\alpha\beta)dz^2}{z^2} -2\alpha\frac{dz d\zeta}{z} \ . \label{nr2}
\end{align}\
We see that up to an overall scaling, the deformation by $\frac{dt dz}{z^3}$ and $\frac{dz d\zeta}{z}$ can be undone by the simple coordinate change. The global structure is slightly changed by the rescaling of variables. Conversely, under certain reasonable assumptions, the non-relativistic conformal background embedded in 2-dimension higher is {\it locally} unique as was shown in \cite{SchaferNameki:2009xr}.

To see this we simply set $\alpha = 0$ and rewrite the metric as
\begin{align}
ds^2_{d+2} = 2\frac{(-1+a^2)dt^2}{z^4} -\frac{2adtdz}{z^3}+\frac{-2dtd\zeta + dx_i^2 +dz^2}{z^2} \ , \label{ans}
\end{align}
where we have rescaled $(t,\zeta) \to (\sqrt{1-a^2}t, \frac{\zeta}{\sqrt{1-a^2}})$ so that $\zeta \simeq \zeta + 2\pi \sqrt{1-a^2} R$. In this coordinate, the non-relativistic conformal transformation are realized in a slightly different manner than in the metric \eqref{nr}:
\begin{align}
(t,\zeta,x_i,z) \to \left(\frac{t}{1+\eta t},\zeta -a\log(1+\eta t)-\frac{\eta}{2}\frac{x_ix_i+z^2}{1+\eta t}, \frac{x_i}{1+\eta t},\frac{z}{1+\eta t} \right) \ .
\end{align}

In the limit $a \to 0$ we recover the original metric \eqref{nr}. On the other hand, in the opposite limit $a \to 1$, our metric is locally diffeomorphic to $AdS_{d+2}$ space (by the simple coordinate transformation $\zeta \to \zeta + \log z$). The latter space was studied in \cite{Goldberger:2008vg}.  One can regard the metric \eqref{ans} as a one-parameter interpolation between the metric studied in \cite{Son:2008ye}\cite{Balasubramanian:2008dm} and that in \cite{Goldberger:2008vg}, where the both claim that the each model is dual to a non-relativistic conformal field theory. They are rather continuously connected.

We see that some components of the curvature tensor are singular near $z \to 0$ as in $a=0$ background just because they are locally diffeomorphic. A test particle will feel the infinite tidal force near the conformal boundary. One particular component of the Riemann tensor, for instance, scales as $R_{\zeta t z t} \propto a(1-a^4)/z^5$, and it appears more singular than in the case $a=0$. Note that all these are just a consequence of the scale invariance except for a proportionality constant, so they are not unexpected. Again one can remove this apparently more singular behavior by the coordinate transformation.

We now move on to the equation of motion that has a solution \eqref{ans}. We first note that the Einstein tensor for the metric \eqref{ans} is given by the sum of ``vacuum energy" $\propto g_{\mu\nu}$ and the ``dust" that has only component in $T_{tt}$:
\begin{align}
T_{\mu\nu} = -\tilde{\Lambda} g_{\mu\nu} - \tilde{E} \delta^{0}_{\mu}\delta^0_\nu g_{00}
\end{align}
 To realize this form of the energy-momentum tensor, let us consider the following Einstein action coupled with a massive vector field (Proca action):\footnote{It is also possible to replace the massive vector field with a particular scalar electro-dynamics in the broken phase as in \cite{Balasubramanian:2008dm}.}
\begin{align}
S = \int d^{d}x_i d\zeta dt dz \sqrt{-g}\left(\frac{1}{2}R -\Lambda -\frac{1}{4} F_{\mu\nu}F^{\mu\nu} - \frac{m^2}{2}A_{\mu}A^{\mu}\right) \ , \label{minv}
\end{align}
where $F_{\mu\nu} = \partial_\mu A_\nu - \partial_\nu A_\mu$. One can show that  $A = A_\mu dx^\mu = -\frac{\sqrt{1-a^2}}{z^2}dt$ solves the Proca equation as well as the Einstein equation, provided
\begin{align}
\Lambda = -\frac{1}{2}(d+1)(d+2) \ , \ \  m^2 = 2(d+2) \ .
\end{align}
In particular, for $a=1$, the metric \eqref{ans} is the solution of the vacuum Einstein equation with cosmological constant. In this case, it is not necessary to introduce the massive vector field.

Finally, we would like to briefly mention Class 2 deformation given by $d\zeta^2$. For simplicity, we turn off Class 3 deformations because they are locally cancelled by a coordinate transformation. 
\begin{align}
ds^2_{d+2} = -2\frac{dt^2}{z^4}+\frac{-2dtd\zeta + dx_i^2 +dz^2}{z^2} + c d\zeta^2 \ . \label{nr3}
\end{align}
Class 2 deformation breaks Galilean invariance as well as non-relativistic conformal invariance, and the Einstein tensor becomes truly anisotropic. One can no longer decompose the Einstein tensor into the vacuum energy and the dust contribution unlike in Class 3 deformation. Explicitly, the Einstein tensor takes the following form:
\begin{align}
G_{tt} &= \frac{6-d(d+1)(1+2c)}{(1+2c)z^4} \ , \  &G_{t\zeta} =- \frac{12c+(d+1)(d+2)(1+2c)}{2(1+2c)z^2} \ , \cr
G_{\zeta\zeta} &= c\frac{-4+4c+ (d+2)(d+3)(1+2c)}{(1+2c)} \ ,  \   &G_{ii} = \frac{4c + (d+1)(d+2)(1+2c)}{2(1+2c)z^2}\  , \cr
G_{zz} &= \frac{-4c+(d+1)(d+2)(1+2c)}{2(1+2c)z^2}\ ,  & \label{mem}
\end{align}
where all the other components are zero.

What kind of matter configuration can realize such energy-momentum tensor? We first note that there exists an energy flow in $\zeta$ direction and the pressure is anisotropic. It can be shown that the massive vector field minimally coupled to gravity as in \eqref{minv} cannot support the energy-momentum tensor \eqref{mem} on shell. It is, nevertheless, possible to introduce further matters/interactions in the action to support the energy-momentum tensor. Since it is not particularly illuminating to do this without specifying the underlying gravitational theory such as the string theory, we briefly mention one bottom-up construction. 

Possible independent tensor structures of the energy-momentum tensor constructed from $A = a_t\frac{dt}{z^2} + a_\zeta d\zeta$ are given by $g_{\mu\nu}, A_{\mu}A_{\nu}$, $F_{[\mu\alpha}F^{\alpha}_{\ \nu]}$ and  $F_{[\mu\alpha}F^{\alpha\beta}F_{\beta\rho}F^{\rho}_{\ \nu]}$ up to the fourth order without higher derivatives. Other combinations are not independent or simply vanish upon symmetrization. One can now construct the desired energy momentum tensor by an appropriate linear combination:
\begin{eqnarray}
T_{\mu\nu} = \tilde{\Lambda} g_{\mu\nu} + g_1 A_{\mu}A_{\nu} + g_2 F_{[\mu\alpha}F^{\alpha}_{\ \nu]} + g_3 F_{[\mu\alpha}F^{\beta\gamma}F_{\gamma\delta}F^{\delta}_{\ \nu]} \ . \label{tens}
\end{eqnarray}
We can show that \eqref{tens} is able to support the energy momentum tensor \eqref{mem} by tuning $a_t$ and $a_\zeta$ (and the coupling constants $g_i$). Note that we have five independent components of the Einstein equation, so the number of unknowns is sufficient.
The equations of motion for $A$ should be fine-tuned as well so that the given $a_t$ and $a_\zeta$ are solutions, which is always possible in principle.

\subsection{Scale invariant deformation from Lifshitz-type background}
So far, in our discussion, we have assumed that the scale invariant deformation preserves the $U(1)$ particle number. Once we relax the condition, there is no reason to retain $\zeta$ direction any more. Indeed, the non-relativistic gravity background that does not have additional $U(1)$ isometry has been discussed in  \cite{Kachru:2008yh}. The metric takes the following form
\begin{align}
ds_{d+1} = -\frac{dt^2}{z^4} + \frac{dx_i^2 + dz^2}{z^2} \ . \label{lf}
\end{align}
The energy momentum tensor is supported by the massive vector field as in \eqref{minv} without $\zeta$ direction, where $m^2 = d^2$, $\Lambda  = -\frac{1+(d+1)^2}{2}$ and $A = \frac{dt}{\sqrt{2}z^2}$.

The metric \eqref{lf} is invariant under
\begin{itemize}
	\item Translations in $(t,x_i)$.
	\item Rotations of $x_i$.
	\item Scale invariance: $ (t,x_i,z) \to (\lambda^2 t , \lambda x_i, \lambda z) $ so the dynamical critical exponent $\mathcal{Z} = 2$.
\end{itemize}
The metric is not invariant under the Galilean boost.

Possible scale invariant deformations are restricted:
\begin{itemize}
	\item Class 1': The deformations that preserve the same symmetry as \eqref{lf} given by $\delta ds^2 = \left\{ \frac{dtdz}{z^3}, \frac{dx_idz}{z^2} \right\}$. They are rather trivial because it can be undone by the coordinate transformation $(t,x_i) \to (t + \alpha z^2, x_i + \beta_i z)$.
	\item Class 2': The deformations that break the rotational invariance as well as time reversal or spatial parity: $\delta ds^2 = \left\{ \frac{dtdx_i}{z^3}\right\}$. This is the genuine deformation of the theory. Our later applications, however, do not  assume such a further symmetry breaking of rotational invariance or parity, so we will not pursue this direction in this paper.
\end{itemize}
Conversely, as studied in \cite{SchaferNameki:2009xr}, the $d+1$ dimensional metric that has the same symmetry as the $d$ dimensional Lifshitz field theory is locally unique.

\subsection{Holographic correlation functions}
Having obtained the gravity background with the right symmetry, we can compute the correlation functions of the corresponding boundary theories by using the same holographic technique employed in AdS/CFT correspondence. For this purpose, firstly, we consider the scalar field propagating in the background \eqref{ans} with the minimal action:
\begin{align}
S = -\int d^dx_i d\zeta dt dz \sqrt{-g}(g^{\mu\nu} \partial_\mu \phi^\dagger \partial_\nu \phi + m_0^2 \phi^\dagger\phi) \ . \label{sa}
\end{align}
As discussed before, $\zeta$ direction is compactified, and we focus on the
mode with $\zeta$ KK momentum $i\partial_\zeta\phi_M = M\phi_M$. With this assumption the field $\phi$ is dual to the boundary operator $O$ which has a definite particle number $M$.

After Fourier transforming in space-time directions, the equation of motion takes the form
\begin{eqnarray}
\partial_z^2 \phi -\frac{1}{z}(d+1-2iMa)\partial_z\phi + \left(2Mw - k^2 - \frac{m^2}{z^2}- \frac{(d+2)iMa}{z^2}\right)\phi = 0 \ ,
\end{eqnarray}
where $m^2 = m_0^2 + (2-a^2)M^2$.
A solution relevant to us is
\begin{eqnarray}
\phi = z^{d/2 + 1 -iMa}K_{\nu}(pz) \ ,
\end{eqnarray}
where $p = \sqrt{k^2-2Mw}$ and $\nu = \sqrt{m^2-M^2a^2 + \frac{(d+2)^2}{4}}$.

The two-point function of primary operators in the field theory is fixed by the non-relativistic invariance \cite{Henkel:1993sg}\footnote{More precisely, the Galilean invariance is sufficient to fix the form of the two-point function.} up to an overall constant and it is given by
\begin{eqnarray}
\langle O(k,w) O^\dagger(-k,-w)\rangle \sim (k^2 - 2Mw)^{\nu} \ ,
\end{eqnarray}
which we can explicitly confirm by the holographic computation.\footnote{This is not a trivial result even if our metric is locally diffeomorphic to the one studied in \cite{Son:2008ye}\cite{Balasubramanian:2008dm}. Recall that the coordinate transformation in the bulk may induce a similar transformation in the boundary theory. For instance, in the usual relativistic AdS/CFT, the correlation functions in the global AdS space are different from those in the Poincar\'e patch, related by the global conformal transformation. The computation here shows that our coordinate change does not induce such a global transformation in the boundary theory.}

In the coordinate space, it is 
\begin{eqnarray}
\langle O(x,t) O^\dagger(0,0) \rangle \sim \theta(t) \frac{1}{|\epsilon^2 t|^{\Delta}} e^{-\frac{iMx^2}{2|t|}} \ , \label{ncprop}
\end{eqnarray}
where we have introduced UV cut-off $\epsilon$ and used the same regularization employed in \cite{Balasubramanian:2008dm}. The scaling dimension $\Delta$ is related to $\nu$ as $\Delta = \frac{d+2}{2} \pm \nu$, where $-$ signature is only possible for $0<\nu<1$ (see \cite{Son:2008ye}). 

Note the specific causal structure of the propagator in the non-relativistic conformal field theory, which is manifested in the step function $\theta(t)$  in \eqref{ncprop}. This causal structure will be relevant in the application to quantum finance in section 3. As we will see, once Galilean invariance is broken, this specific causal structure could be lost.

Let us now consider the Galilean violating Class 2 deformation \eqref{nr3}. Again for simplicity we turn off Class 3 deformation. The equation of motion for the scalar field as in \eqref{sa} becomes
\begin{eqnarray}
\partial^2_z \phi -\frac{1}{z}(d+1)\partial_z\phi + \left(\frac{c}{1+2c} w^2 z^2 -k^2 + \frac{2Mw}{1+2c}- \frac{m^2}{z^2}\right)\phi = 0 \ , \label{see}
\end{eqnarray}
where $m^2 = m_0^2 + \frac{2M^2}{1+2c}$. 
The solution of the scalar equation of motion relevant to our study is given by
\begin{align}
&\phi = e^{\frac{\sqrt{c} w z^2}{i2\sqrt{2c+1}}} z^{\frac{d}{2} + 1 + \nu}  \times \cr
&U\left( \frac{(1+2c)k^2 +2(-M + i\sqrt{c(2c+1)}(1+\nu))w}{4i\sqrt{c(2c+1)}w}, 1 + \nu; -\frac{\sqrt{c} w z^2}{i\sqrt{1+2c}}\right) \ ,
\end{align}
where $U(a,b;x)$ is the confluent Hypergeometric function (we refer appendix C for details), and we have introduced $\nu = \sqrt{m^2+ \frac{(d+2)^2}{4}}$ as before (with $a=0$). Note that in this expression, $c \to 0$ is singular and it requires a careful treatment.

The bulk boundary propagator $G(z,w,k)$ satisfies the same equation of motion \eqref{see} with the boundary condition $G(\epsilon,w,k) = 1$. 
Subsequently, the boundary two-point function is computed as
\begin{align} 
& \langle O(k,w) O^\dagger (-k,-w) \rangle \cr
&= G(\epsilon,k,w) \sqrt{g} g^{zz} \partial_z G(\epsilon,k,w) \cr
& = \mathcal{C} w^{\nu} \frac{\Gamma(-\nu) \Gamma\left(\frac{(1+2c)k^2 +2(-M + i\sqrt{c(2c+1)}(1+\nu))w}{4i\sqrt{c(2c+1)}w}\right)}{\Gamma(\nu) \Gamma\left(\frac{(1+2c)k^2 +2(-M + i\sqrt{c(2c+1)}(1-\nu))w}{4i\sqrt{c(2c+1)}w}\right)} + \text{ultra local terms} \ , \label{holc2}
\end{align}
where we have omitted the ultra local contribution (= contact terms) in the boundary propagator. $\mathcal{C}$ is a normalization constant independent of $w$ and $k$.

It is interesting to note that the boundary two-point function is same as that for the Lifshitz-type background if one sets $M=0$ up to a straightforward rescaling of parameters.\footnote{It is important to note that the limits $M\to 0$ and $ c \to 0$ do not commute: if one first took $c\to 0$, the dynamics in $M=0$ sector would be completely frozen as shown in \cite{Nakayama:2009ed}.} The latter has been computed as \cite{Kachru:2008yh}
\begin{eqnarray}
\langle O(k,w) O^\dagger (-k,-w) \rangle = \mathcal{C}w^{\nu} \frac{\Gamma(-\nu) \Gamma\left(\frac{k^2 +2i(1+\nu)w}{4iw}\right)}{\Gamma(\nu) \Gamma\left(\frac{k^2 +2i(1-\nu)w}{4iw}\right)} + \text{ultra local terms} \ . \label{kpr}
\end{eqnarray}
A priori, this is not expected because the breaking of the Galilean invariance could introduce any function of $f(k^2/w)$ in the boundary two-point function. The role of non-zero $M$ simply shifts the momentum $k^2 \to k^2 -\frac{2Mw}{1+2c}$.

\subsection{Deviation from $\mathcal{Z}= 2$ and Galilean invariance}
We can generalized our construction by relaxing the dynamical critical exponent $\mathcal{Z}=2$. Although the non-relativistic conformal symmetry is broken, we are still able to preserve the Galilean invariance. The geometry corresponding to Galilean invariant field theory with $\mathcal{Z} \neq 2$ is given by
\begin{eqnarray}
ds_{d+2} = -\frac{2dt^2}{z^{2\mathcal{Z}}} + \frac{-2d\zeta dt + dx_i^2 + dz^2}{z^2} \ .
\end{eqnarray}
Without breaking the Galilean invariance, one can introduce an analogue of Class 3 deformation by $\delta ds^2 = \left\{ \frac{d\zeta dz}{z}, \frac{dt dz} {z^{\mathcal{Z}+1}}\right\}$. Note that $\zeta$ now possesses a non-trivial scaling $\zeta \to r^{2-\mathcal{Z}}\zeta$. 

The two-point function of primary operators in this background is fixed by the Galilean invariance. In the momentum space, it is given by
\begin{eqnarray}
\langle O(w,k) O^\dagger(-w,-k) \rangle \propto (k^2-2Mw)^\nu \ . \label{2ptf}
\end{eqnarray}
At first sight, this seems to be inconsistent with the dynamical critical exponent $\mathcal{Z} \neq 2$. However, one should be reminded that the particle number $M$ also scales under the scaling symmetry when $\mathcal{Z} \neq 2$: $i[D,M] = (2-\mathcal{Z})M$ so that \eqref{2ptf} transforms with weight $2\nu$ under the dilatation. Note that because of this non-trivial scaling, $\zeta$ direction cannot be compactified. It may be rather interpreted as $d+1$ dimensional field theory with additional coordinate $\zeta$.

With this difficulty, we have to abandon the Galilean invariance to compactify $\zeta$ direction. The general metric ansatz would be
\begin{align}
ds^2 = -2\frac{dt^2}{z^{2\mathcal{Z}}} + \frac{-2d\zeta dt}{z^{\mathcal{Z}}} + \frac{dx_i^2 + dz^2}{z^2} + cd\zeta^2 \ ,
\end{align}
where $\zeta$ is invariant under the dilation so that we can compactify $\zeta \sim \zeta + 2R$.

In order to study the correlation functions, we introduce a minimally coupled scalar field $\phi$ as in \eqref{sa}. The equation of motion is given by
\begin{eqnarray}
\partial_z^2 \phi - \frac{1}{z}(d+1) \partial_z \phi + \left(\frac{cz^{2\mathcal{Z}-2}}{1+2c} w^2 - k^2 +\frac{2Mw}{1+2c}z^{\mathcal{Z}-2} - \frac{m^2}{z^2}\right)\phi = 0 \ , \label{unsolve}
\end{eqnarray}
where $m^2 = m_0^2 + 2\frac{M^2}{1+2c}$.
As in $\mathcal{Z} =2$, by setting $M=0$, we obtain the same scalar equation of motion as in the Lifshitz type background studied in \cite{Kachru:2008yh} (up to rescaling of variables).

In order to compute the boundary two-point functions, we have to solve the equation \eqref{unsolve} to obtain the bulk-boundary propagator. Unfortunately, we have been unable to find simple analytical solutions except for $\mathcal{Z}=4$ and $c=0$. In this special case, we find that the field profile is
\begin{eqnarray}
\phi = e^{-\frac{i\sqrt{Mw}}{\sqrt{2}}z^2} z^{\frac{d+2}{2} + \nu} U\left(\frac{\sqrt{2}k^2}{i\sqrt{Mw}} +\frac{1}{2}(1 + \nu) , 1+\nu; i\sqrt{2Mw} z^2\right) \ ,
\end{eqnarray}
where $\nu = \sqrt{m^2 + \frac{(d+2)^2}{4}}$. Accordingly, the two-point function is computed as
\begin{eqnarray}
\langle O(k,w) O^\dagger (-k,-w) \rangle = \mathcal{C}w^{\nu} \frac{\Gamma(-\nu) \Gamma\left(\frac{\sqrt{2}k^2}{i\sqrt{Mw}} + \frac{1+\nu}{2} \right)}{\Gamma(\nu) \Gamma\left(\frac{\sqrt{2}k^2}{i\sqrt{Mw}} + \frac{1-\nu}{2} \right)} + \text{ultra local terms} \ .
\end{eqnarray}
 A similar observation has been made in \cite{SekharPal:2008uy}.

\section{Applications}
Given the gravity solution, we discuss possible applications to the strongly coupled dual system in this section. We pick up two particular subjects, i.e. the scaling regime of the Reggeon field theory and the non-linear quantum finance. While our gravity solution might have more conventional applications such as in condensed matter physics, we choose rather ``non-standard" applications so as to expand the realm of the ``gauge/gravity" correspondence as discussed in Introduction. We hope our discussions will stimulate the further study on these yet unexplored subjects.

\subsection{Reggeon field theory}
Non-relativistic conformal invariance was first discovered as a symmetry of the free Sch\"odinger equation \cite{Hagen:1972pd}\cite{Niederer:1972zz}. The symmetry itself, however, could appear as an effective action of some quasi-particles much like the relativistic dispersion relation and the emergent ``Poincar\'e invariance" that can be realized in condensed matter systems. 

One of such realizations is the physics of Reggeon in the high energy scattering regime. It has been shown that in the large $\mathbf{s}$ limit with fixed momentum transfer $\mathbf{t}$,\footnote{We denote the Mandelstam variables by boldface such as $\mathbf{s}$ and $\mathbf{t}$.} the Reggeon exchange contributing to the partial wave expansion of the scattering amplitude is described by the so-called Reggeon field theory (see e.g. \cite{Abarbanel:1975me}\cite{Moshe:1977fe} for reviews). The Reggeon field theory is a $(1+2)$ dimensional field theory living in $t \sim \log\mathbf{s}$ and transverse $x_{\perp} = (x_1,x_2)$ plane. The ``energy" $E$ of the Reggeon field theory is identified with the complex angular momentum $1-J$ of the underlying relativistic field theory (QCD).

The non-interacting Reggeon field theory has a $(1+2)$ dimensional dispersion relation
\begin{eqnarray}
 E = 1-J = \alpha' k^2 \ ,
\end{eqnarray}
where $\alpha'$ is the Regge slope. The corresponding field theory can be formulated by the ``Schr\"odinger" action
\begin{eqnarray}
S = \int dt dx^2\left( i \psi^\dagger \partial_t \psi - \alpha' (\partial_i \psi^\dagger \partial_i \psi) - V(\psi) \right) \ .
\end{eqnarray}
Note that the free part without the potential $V(\psi)$ has the full symmetry of the non-relativistic conformal invariance.

We would like to understand the IR limit of the Reggeon field theory, where the theory is supposed to show scaling invariance. Our hope is that with the non-trivial potential $V(\psi)$, the IR limit may show non-trivial strongly coupled fixed point that could be studied by the gravity dual discussed in section 2. As we have seen in section 2, the structure of the gravity dual is rather robust, so we hope that the prediction from the gravity dual is universal.

The properties of the IR fixed point theory depend on the symmetry assumption about the potential $V(\psi)$. In the literature \cite{Gribov:1968fc}, it has been suggested that, when the Reggeon is identified with the Pomeron, which has the vacuum quantum number, the lowest order interaction is given by the cubic coupling with a non-Hermitian coefficient:
\begin{eqnarray}
V_{3P}(\psi) = ir (\psi^2 \psi^\dagger + \psi^{\dagger2} \psi) \ , \label{trip}
\end{eqnarray}
which describes the three Pomeron interaction with intensity $r$. The non-trivial zero of the beta function for $r$ has been found within the $\epsilon = 4-d$ expansion. Note that the cubic coupling breaks particle number conservation and it naturally generates a non-zero mass term $V_{mass} = \psi^\dagger\psi$ as quantum corrections, so in order to obtain a non-trivial scaling regime in IR, we have to fine-tune the bare mass parameter (much like the Wilson-Fisher fixed point of $\lambda \phi^4$ theory in $4-\epsilon$ dimension).

Alternatively, one could imagine a hypothetical world where the Reggeon has a conserved particle number. In our real word, the Pomeron cannot have such a conserved quantum number, but we might expect that higher Reggeon might be described by such a conserved quasi-particle in a certain parameter regime.
In that case, the lowest interaction would be
\begin{align}
V_{4P}(\psi) = \lambda |\psi|^4 \ .
\end{align}
The interaction preserves the non-relativistic conformal invariance once we assume the existence of a non-trivial fixed point. See appendix A for more discussions on the conformal invariance at the one-loop order.
The gravity dual that has the non-relativistic conformal invariance has been discussed in section 2.1. The Reggeon propagator is given by
\begin{eqnarray}
G(k,E) = \frac{\mathcal{C}}{(E-\alpha' k^2)^\nu} \ ,
\end{eqnarray}
whose form is fixed by the Galilean invariance.

Now let us go back to the originally proposed cubic Pomeron interaction. Since the non-relativistic conformal invariance as well as the particle number conservation is broken, we expect the gravity dual is close to that studied in section 2.2, if any. Can we understand the physics of the Reggeon field theory from the Lifshitz type gravity background or its 1-dimensional higher counterpart \eqref{nr3}? 

The key ingredient to answer this question is to understand the dispersion relation from the predicted two-point function \eqref{holc2} from the gravity. We can easily see that in the physical parameter region, there are no poles in \eqref{holc2} or \eqref{kpr} by noting $\Gamma(z)$ has only single poles at negative integers (including zero), and $\Gamma(z)$ does not have zero anywhere in $z$ plane. Thus we see that the gravity computation would not predict any quasi-particle excitation for the Reggeon field theory.

One might wonder whether the prediction makes sense as a Reggeon field theory. We could circumvent the problem by doing a specific analytic continuation in \eqref{holc2} to use a negative value of $-\frac{1}{2} \le c \le 0$. Then, the Gamma function yields poles at $w = \alpha'_{eff} k^2$, where the effective Regge slope is given by
\begin{align}
 \alpha'_{eff} = \frac{1+2c}{2M - i\sqrt{c(2c+1)}(1+\nu)-4iN\sqrt{c(2c+1)}} \  \label{effp}
\end{align}
with a negative integer $N = 0, -1, -2, \cdots$. Accordingly, the two-point function near the pole scales as\footnote{It is interesting to note that the quasi-particle spectrum is impossible in non-trivial relativistic conformal field theories; see e.g. \cite{Georgi:2007ek} for recent applications to (un)particle physics. The non-relativistic scale invariance can accommodate such a quasi-particle spectrum density in the two-point function.}
\begin{eqnarray}
G(k,E) = \frac{\mathcal{C}E^{-\nu+1}}{E-\alpha'_{eff} k^2} \ . \label{reggep}
\end{eqnarray}
The form is also predicted from the $\epsilon$ expansion of the Reggeon field theory (by assuming $\mathcal{Z}=2$: see \cite{Moshe:1977fe} and references therein). We note that $\Gamma$ function yields infinitely many poles with different effective Regge slope $\alpha'_{eff}$. The residues of the poles have alternating signatures, so the higher poles may not be physical. $m=0$ gives the largest Regge slope.

From the gravity viewpoint, the reason why we have to take negative $c$ is not obvious. The $\zeta$ direction is a  closed time-like curve. On the other hand, the Reggeon field theory by itself is non-unitary because the interaction is anti-Hermitian, so one might not exclude the possibility that the gravity dual could be a non-unitary theory as indicated by the negative $c$.

Finally, we would like to study the deviation from $\mathcal{Z} =2$. While the tree level action of the Reggeon field theory is compatible with the $\mathcal{Z}=2$ scaling, there is no protection of the dynamical critical exponent $\mathcal{Z}$ from the renormalization once we break the Galilean invariance (and particle number conservation). Indeed, the two-loop $\epsilon$ expansion predicts that the dynamical critical exponent would be \cite{Baker:1974yr}\cite{Bronzan:1974jd}
\begin{eqnarray}
\frac{2}{\mathcal{Z}} = 1 + \frac{\epsilon}{24} + \frac{\epsilon^2}{144}\left(\frac{59}{24}\log \frac{4}{3} + \frac{79}{48}\right) \ .
\end{eqnarray}
Similarly, to the first order in $\epsilon$, the two-point function scales as
\begin{eqnarray}
G(k,E) = \frac{\mathcal{C}E^{\gamma}}{(E-\alpha' k^2E^{\xi})} \ ,
\end{eqnarray}
where $\mathcal{Z} = \frac{2}{1-\xi} = 1.74$, and the anomalous dimension $\gamma$ is given by
\begin{eqnarray}
\gamma = -\frac{\epsilon}{12}-\frac{\epsilon^2}{144}\left(\frac{161}{12}\log\frac{4}{3}+\frac{37}{24}\right)
\end{eqnarray}
In particular, $\gamma = -0.32$ for $d= \epsilon = 2$. 

Since the relation between the underlying relativistic field theory and the effective Reggeon field theory is quite non-trivial, the unitarity bound (e.g. Froissart bound and the condition that the elastic cross section is less than the total cross section: $\sigma_{el} \le \sigma_{tot}$) of the former dictates a non-trivial constraint on the Reggeon field theory. The unitarity bound can be translated into $-\gamma \le \frac{2}{\mathcal{Z}} \le 2$ for anomalous dimensions of Reggeon field theory (see \cite{Abarbanel:1975me}\cite{Moshe:1977fe} and references therein). The bound $\frac{2}{\mathcal{Z}}\le 2$ may have a simple interpretation from the gravity. As shown in \cite{Kachru:2008yh} only in this case, the energy-momentum tensor can be supported by the flux of massive vector field. Physically, the dual theory cannot show critical speeding up faster than $\mathcal{Z}=1$.  The other bound $-\gamma \le \frac{2}{\mathcal{Z}}$ is equivalent to the condition for the conformal dimension: $\Delta \ge \frac{d}{2}-2$. When the theory is conformally invariant, the condition is weaker than the unitarity bound $\Delta \ge \frac{d}{2}$. In this way, the unitarity bound for the underlying theory is related to the physical requirement of the dual gravitational theory.

The corresponding dual gravity has been described in section 2.4. Unfortunately, the gravity solution cannot predict the value of $\mathcal{Z}\ge 1$. Only we could do is to study the correlation functions with a given dynamical critical exponent $\mathcal{Z}$. In particular, it would be interesting to study the causal structure and dispersion relation of the two-point function gravity as we have done in $\mathcal{Z} =2$ above. We defer this study because the analytic expression for the gravity two-point function is unavailable, and we would have to resort to the numerical computation. In later section 3.4 we will come back to the issue of non-trivial dynamical critical exponent and its relation to fractal geometry in the context of non-linear quantum finance.

\subsection{Linear quantum finance}
Before we discuss the application of our holographic computation to the non-linear quantum finance, we briefly review the classical Black-Scholes-Merton model of option pricing \cite{BS}\cite{M} from the (supersymmetric) path integral viewpoint.\footnote{We refer \cite{quant1}\cite{quant2}\cite{quant3} for reviews of quantum physics approach to finance with the usage of path integral.} Since the computation is reduced to a linear differential equation i.e. ``Schr\"odinger equation", we call it as linear quantum finance.

Our starting point is the generalized Langevin equation:
\begin{eqnarray}
\frac{\partial X(t)}{\partial t} = \frac{\partial W(X)}{\partial X} + \sigma\eta(t) \ , \label{gle}
\end{eqnarray}
where $\eta(t)$ is a Gaussian random noise that satisfies 
\begin{align}
\langle \eta(t) \eta(0) \rangle = \delta(t) \ . \label{GR}
\end{align}
In the Black-Scholes-Merton model, the ``superpotential" $W(X)$ is simply given by a linear function $W(X) = \mu X$, where $X(t)$ is the logarithm of the risky asset price $S(t)$ (i.e. $S(t) = e^{X(t)}$), but for a while we keep the superpotential more general.

Suppose we would like to compute the expectation value of a certain quantity $F(X)$ averaged over the random paths satisfying \eqref{gle}. By introducing the Gaussian measure for the random noise $\eta(t)$ to realize \eqref{GR}, one can express it in the form of the path integral as
\begin{align}
\langle F(X) \rangle = \int [\mathcal{D} \eta][\mathcal{D}X] F(X)  J(X) \delta\left(\frac{\partial X}{\partial t} -\frac{\partial W}{\partial X} - \sigma\eta\right)\exp\left(-\int dt \frac{\eta^2}{2}\right) \ .
\end{align}
The Jacobian appearing in the measure is the usual Fadeev-Popov factor associated with the delta functional constraint to impose \eqref{gle}, and it can be exponentiated to a form of action by introducing Grassmann fields $\bar{\psi}(t)$ and $\psi(t)$ as
\begin{align}
J(X) =\int [\mathcal{D}\psi][\mathcal{D}{\bar{\psi}}] \exp\left(-\frac{1}{\sigma^2}\int dt \bar{\psi}\partial_t \psi -  \frac{\partial^2 W}{\partial X^2}\bar{\psi}\psi \right) \ . \label{jacob}
\end{align}

On the other hand, the Gaussian noise $\eta$ can be removed in the path integral by using the delta functional constraint, so the Bosonic action becomes
\begin{align}
S_{bos} = \int dt \frac{1}{2\sigma^2}(\partial_t X - \partial_X W)^2 \ . \label{bosa}
\end{align}
We see that the total action is nothing but that of the supersymmetric quantum mechanics in the Euclidean signature \cite{Parisi:1982ud} upon partial integration of the cross term $\partial_tX \partial_XW = \partial_t W$ in the Bosonic action.

The Black-Scholes-Merton model specifies the superpotential $W = \mu X$ and discusses the geometric Brownian motion. Since the superpotential is linear, the fermionic part completely decouples, so we can safely neglect the Jacobian factor \eqref{jacob} in the following. To interpret the parameters in the action, we note that the no-arbitrage assumption requires $\mu = r -\frac{1}{2}\sigma^2$, where $r$ is the risk free interest rate (say, bank interest rate), and $\sigma$ is the volatility of the risky asset (say, stock).

Let us now compute a present value of a European call option as the simplest application of the Black-Scholes-Merton model. For this, we simply compute the (discounted) expectation value for 
\begin{align}
C(X^*,t^*) = (e^{X^*} -E)\theta(e^{X^*} - E) \ , \label{price}
\end{align}
where $E$ is the exercise price of the call option, and $X^*$ is the value of the asset at the maturity time $t^*$.\footnote{A European option is a contract that entitles one to buy a share of stock at time $t^*$ at a fixed price $E$. He will gain $S(t^*)-E$ when the stock price $S(t^*) = e^{X^*}$ is higher than $E$. On the other hand, he does not lose anything when the stock price is lower than $E$ because he does not have to buy the stock. This is the origin of the step function in \eqref{price}.} We also impose the boundary condition $X(t_0) = X_0 = \log S_0$. Explicitly, the discounted European option price is given by\footnote{Because of the risk free interest rate $r$, the value of the option is naturally deflated by the factor $e^{-r (t^*-t_0)}$.}
\begin{align}
&C(X_0,t_0) \cr
&= e^{-r (t^*-t_0)} \int_{\log E}^{\infty} dX^* \int_{X(t_0)=X_0}^{X(t^*)=X^*} [\mathcal{D}X] (e^{X^*} -E)\exp\left(-\int dt \frac{1}{2\sigma^2}(\partial_t X - \mu)^2 \right) \ .
\end{align}

The Gaussian path integral for $X(t)$ is readily performed as
\begin{eqnarray}
C(X_0,t_0) = \int_{\log E}^{\infty} dX^* (e^{X^*}-E) \mathcal{G}(X_0-X^*,t_0-t^*) \ , \label{bsi}
\end{eqnarray}
where the propagator $\mathcal{G}(X_0-X^*,t_0-t^*)$ is just given by
\begin{eqnarray}
\mathcal{G}(X_0-X^*,t_0-t^*) = \frac{e^{-r(t^*-t_0)} }{\sqrt{2\pi \sigma^2(t^*-t_0)}} \exp\left(-\frac{(X_0-X^* + \mu (t^*-t_0))^2}{2\sigma^2 (t^*-t_0)} \right) \ ,
\end{eqnarray}
which is nothing but a free particle propagator (with a constant external vector potential) corresponding to the Bosonic action \eqref{bosa}. One may easily perform the remaining integral \eqref{bsi} to obtain the celebrated Black-Scholes formula for the European call option:
\begin{align}
C(X_0,t_0) = S_0N(d_1) -E e^{-r(t^*-t_0)}N(d_2) \ ,
\end{align}
where
\begin{align}
d_1 &= \frac{\log(S_0/E) + (r+\frac{\sigma^2}{2})(t^*-t_0)}{\sigma\sqrt{t^*-t_0}} \cr
d_2 &= d_1 -\sigma\sqrt{t^*-t_0}
\end{align}
and $N(x) = \frac{1}{\sqrt{2\pi}}\int_{-\infty}^x e^{-\frac{t^2}{2}} dt$.

By construction, we can regard $C(X_0,t_0)$ as the ``wavefunction". Indeed, it satisfies the ``Schr\"odinger equation" (or Kolmogorov backward equation):
\begin{eqnarray}
\frac{\partial C}{\partial t} = r C -\mu \frac{\partial C}{\partial X} - \frac{1}{2} \sigma^2 \frac{\partial^2 C}{\partial X^2} \ . \label{BSE}
\end{eqnarray}
Unconventional terms $rC$ and $\mu \frac{\partial C}{\partial X}$ can be absorbed by the change of variables $C(X,t)\to e^{rt}C(X-\mu t, t)$ to directly compare with the Wick rotated Schr\"odinger equation. 
Note that with this change of variable, the propagator used in the Black-Scholes-Merton model is simply the Wick rotated version of the Schr\"odinger propagator obtained in the gravity approach in \eqref{ncprop}. The equation \eqref{BSE} is known as the Black-Scholes equation in finance.

The propagator can be obtained either in the first quantized formulation presented here or in the second quantized formulation by quantizing the Schr\"odinger action, both of which are equally valid in the linear quantum finance. We use the second quantized method in the following subsections to discuss the non-linear quantum finance.

\subsection{Toward non-linear quantum finance}
As a first order approximation to the real market, the Black-Scholes-Merton model has been quite successful both theoretically and experimentally. Theoretically, the Black-Scholes-Merton model has an advantage that it possesses equivalent Martingale measure, which fits with the efficient market hypothesis with no arbitrage.\footnote{It roughly means that nobody cannot consistently beat the market no matter how smart he is.}

There are several problems, however:
\begin{itemize}
	\item The fluctuation in the real market is not Gaussian unlike \eqref{GR}. It does not decay as fast as Gaussian (fat tail problem), and the history shows that the catastrophic loss (gain) is more likely than the Gaussian model.\footnote{There is great evidence on this point; see e.g. \cite{quant3}\cite{nr}\cite{Man} and references therein.}

	\item Furthermore, people argue that there is a slight non-trivial time dependence in the market performance. The non-equal time auto-correlation is slightly positive when the time scale is small (i.e. the companies winning continues to win for a while: winner has a momentum). On the other hand, the correlation is slightly negative when the time scale is large (i.e. no company cannot consistently dominate the market: the trend will shift).\footnote{This observation is controversial once we subtracted the underlying uptrend of the whole market. See e.g. \cite{nr} and references therein.} Also some people believe the usage of market cycle.

	\item The Gaussian hypothesis neglects possible non-trivial higher-point correlations (or cumulant) in the market. The non-trivial higher point correlation functions clearly indicate the underlying non-linear nature of the market. People ``hope" that by using non-linear analysis, they may be able to beat the market because the non-linearity might break the efficient market hypothesis.

\end{itemize}

There are many models to account for these issues, but we would like to focus on a particular attempt suggested in \cite{quant2} to use a non-linear quantum field theory because we may be able to tackle the problem by using our gravitational dual realization. The idea is inspired by the appearance of the effective Schr\"odinger equation in the Black-Scholes-Merton model and its analogy to the Reggeon field theory. We take the second quantized approach and regard the particle propagator as
\begin{eqnarray}
\langle \phi(x_0,t_0) \phi^\dagger(x,t) \rangle = \int [\mathcal{D}\phi] \phi(x_0.t_0) \phi^\dagger(x,t) \exp(-S_0[\phi]) \ ,
\end{eqnarray}
where $S_0[\phi] = \int dtdx \left(-\phi^\dagger \partial_t\phi + \frac{\sigma^2}{2}\partial_x\phi^\dagger \partial_x\phi \right)$ is the free Euclidean Sch\"odinger action.

The proposal is that we replace the free Sch\"odinger action by the interacting action 
\begin{eqnarray}
S[\phi] = \int dtdx \left(-\phi^\dagger \partial_t\phi + \frac{\sigma^2}{2}\partial_x\phi^\dagger \partial_x\phi + V(\phi) \right) \ . \label{nlsa}
\end{eqnarray}
In the short range, where $\delta X$ and $\delta t$ are small (i.e. in the UV limit), it is reduced to the original Black-Scholes-Merton model, while when $\delta X$ or $\delta t$ are large (i.e. in the IR limit), the non-linearity will be of importance and the theory may show a non-trivial scaling regime as in the Reggeon field theory. The formulation is quite analogous to the Reggeon field theory except that our ``space time" is $(1+1)$ dimension (or $d=1$) rather than $(1+2)$ dimension (or $d=2$) in Reggeon field theory.

The prescription is simply to replace the free particle propagator in the option pricing formula by the interacting propagator obtained from the non-linear action \eqref{nlsa}. We keep the form of the Schr\"odinger action by performing the replacement $\mathcal{G}(X,t)\to e^{rt}\mathcal{G}(X-\mu t, t)$ at the end of the computation to incorporate the drift and deflation because the Schr\"odinger action possess a manifest enhanced symmetry (non-relativistic conformal symmetry) and it is a  good starting point to discuss its gravitational dual.

We focus on the IR limit of the non-linear theory \eqref{nlsa} so that we may expect to learn the structure of the correlation functions from the gravity dual with the scaling symmetry. As in the Reggeon field theory, the IR structure will depend on the symmetry assumption of the potential $V(\phi)$. The potential could depend on higher derivatives, while these higher derivative terms are always perturbatively irrelevant around the free field fixed point.

Let us first consider the case where the potential $V(\phi)$ preserves the Galilean invariance. Slightly weaker assumption that the theory is invariant under the particle number  leads to the same constraint on the potential as long as we discard any higher derivative interactions. The gravity dual corresponding to this assumption is the non-relativistic conformal invariant background studied in section 2.1. The non-relativistic conformal invariance fixes the form of the propagator. In our context, the propagator should read
\begin{eqnarray}
\mathcal{G}(x,t) = \theta(t) \frac{e^{-rt}}{|2\pi\sigma^2 t|^{\Delta}} e^{-\frac{(x-\mu t)^2}{2\sigma^2 t}} \ 
\end{eqnarray}
from section 2.4.
The only non-trivial parameter is the scaling dimension $\Delta$. Note that the exponential damping is not alleviated and still be present due to the 
non-relativistic conformal invariance. The option-pricing formula would be slightly modified by replacing $\frac{1}{\sqrt{t}}$ factor by $\frac{1}{|t|^{\Delta}}$.

Now, in more general situations, we do not impose the Galilean invariance (e.g. broken by higher derivative interactions), or we even do not expect particle number conservation (e.g. by introducing a cubic coupling as in Reggeon field theory).\footnote{The particle number conservation may be important to understand the conservation of probability in quantum mechanics, while the situation and its necessity is less obvious in the quantum finance. } The gravity dual for such deformations has been studied in section 2.1 and 2.2. 

We have studied the corresponding propagator in section 2.4. The most important feature of the two-point function studied there is its causal structure (see section 3.1). They do not have any quasi-particle dispersion relation as long as $c>0$. The Green function carries information both forward in time and backward in time. This may or may not be a problem for the quantum financial interpretation because we would like to estimate the future option price from the available data at present. Analytic continuation to $c<0$ might be relevant in the present case as in the Reggeon field theory. In that regime, near one pole, the causal structure of the Green function is governed by the particular Regge pole; see \eqref{reggep}. 

Each pole would contribute to the Green function as
\begin{eqnarray}
\sim \theta(t) \left(\frac{x-\mu t}{t}\right)^{2(-\nu+1)}\frac{e^{-rt}}{\sqrt{2\pi\sigma_{eff}^2 t}} e^{-\frac{(x-\mu t)^2}{2\sigma_{eff}^2 t}} \ 
\end{eqnarray}
for large $t$ with $x/t$ fixed as long as the poles are well separated. Here $\sigma_{eff}$ depends on the pole location, which is related to the effective Regge slopes in the gravity dual description of the Reggeon field theory (c.f. \eqref{effp}). Thus, the option price behaves as if it were composed of multiple securities with different effective volatilities. The appearance of the multiple effective volatilities is of great significance in the real market and has been investigated as a first step to improve the Black-Scholes-Merton model (e.g. GARCH model). It may also be related to the multi-fractal analysis of the market pursued in \cite{Man}.

On the other hand, at a specific parameter region of the scale invariant but non-conformal gravity background such as Lifshitz-type background, long distance (fixed time) propagators can be power-like without any exponential tail by tuning the conformal dimension (see \cite{Kachru:2008yh} for a detailed demonstration).
Explicitly, the two-point function of operator $O$ whose scaling dimension is $4$ has been shown to behave as
\begin{eqnarray}
\langle O(t,x) O^\dagger (0,0) \rangle = \frac{\mathcal{C}}{|x|^8} \  
\end{eqnarray}
in $|x| \to \infty$ limit. To obtain this result, we cannot use the formula \eqref{kpr} because the separation between the analytic part and the non-analytic part becomes more complicated than the derivation of \eqref{kpr} when the scaling dimension is an integer. Similarly, whenever $\nu$ is a positive integer, the power-like behavior can appear.
 This also holds in the $M=0$ sector of the deformed theory studied in section 2.1.  The emergent power-like decay is quite promising in the application of non-linear quantum finance to understand the fat tail problem mentioned above.

Finally, once we break the Galilean invariance or particle number conservation, there is no reason why the dynamical critical exponent $\mathcal{Z}$ remains to be 2. Indeed, the non-trivial dynamical critical exponent is proposed in finance as well. As we have discussed in section 3.1, this is also expected in $\epsilon$ expansion of Reggeon field theory with no particle number conservation. We will further investigate this issue in the next subsection.

\subsection{Deviation from $\mathcal{Z}= 2$: Hurst exponent}
We have seen that in Reggeon field theory with cubic non-Hermitian interaction,  the $\epsilon$ expansion around $d=4$ leads to a non-trivial dynamical critical exponent $\mathcal{Z}\neq 2$. This is not totally unexpected because the violation of the non-relativistic conformal invariance induced by the violation of the Galilean invariance does not protect the dynamical critical exponent. 

The concept of dynamical critical exponent was first introduced in finance by Mandelbrot (see \cite{Man} and references therein), and it is called a Hurst exponent. Let us consider a random variable $Q(t)$. When the random variable shows a scaling auto-correlation function
\begin{eqnarray}
\langle |Q(t+T) - Q(t)|^q\rangle = C_qT^{q\mathcal{H}} \label{Hurstd}
\end{eqnarray}
for $q>-1$ with a constant $C_q$, we say that the random variable $Q(t)$ has a Hurst exponent $\mathcal{H}$. Here in order to define $\mathcal{H}$, we have assumed that the underlying distribution for $Q(t)$ has stationary increment so that \eqref{Hurstd} does not depend on $t$. Note that $\mathcal{H}$ could depend on $q$ due to the anomalous dimension of composite operators when the theory is highly non-linear.

For example, let us consider the case $Q(t) = B(t) = \int_0^t \eta(s) ds$ with the Gaussian random noise $\eta(t)$ satisfying \eqref{GR}. $B(t)$ is the Brownian motion, and it is easy to see $\mathcal{H}=\frac{1}{2}$. Similarly fractional Brownian motion
\begin{eqnarray}
Q(t) = B^\mathcal{H}(t) = \int_{-\infty}^0\left( (t-s)^{\mathcal{H}-\frac{1}{2}} -(-s)^{\mathcal{H}-\frac{1}{2}}\right) \eta(s) ds + \int_0^t(t-s)^{\mathcal{H}-\frac{1}{2}} \eta(s) ds \ 
\end{eqnarray}
has a non-trivial Hurst exponent $\mathcal{H}\neq \frac{1}{2}$ which appeared in its definition. The Brownian motion with $\mathcal{H}=\frac{1}{2}$ is Martingale, while the fractional Brownian motion is not Martingale and it shows a memory effect. The Hurst exponent is related to the fractal dimension of the sample path of the random process by $d = 2-\mathcal{H}$.

Schematically one could write \eqref{Hurstd} as $\delta X = (\delta t)^{\mathcal{H}}$. If we compare this definition with our discussions of the non-linear quantum finance, we can identify $\mathcal{H}= \frac{1}{\mathcal{Z}}$ from the dispersion, or scaling relation $E=k^{\mathcal{Z}}$, where $\mathcal{Z}$ is the dynamical critical exponent. Here, it is important to carefully subtract the drift contribution in order to satisfy the stationary condition. The failure to do this in market data leads to a spurious Hurst exponent.
 We recall that the Gaussian assumption leads to $\mathcal{H}=\frac{1}{2}$ and this is the value mostly studied in previous sections in particular with the non-relativistic conformal invariance $(\mathcal{Z}=2)$, which is manifest after the change of variables  $C(X,t)\to e^{rt}C(X-\mu t, t)$. Our non-linear action \eqref{nlsa} has the obvious time-translation invariance.
 Historical analysis of market data suggests that $\mathcal{H}$ is slightly above $\frac{1}{2}$, but not significantly larger \cite{nr}.

An interpretation of the Hurst exponent in finance is that when $\frac{1}{2}<\mathcal{H}<1$, the security is persistent in memory, and when $0<\mathcal{H}<\frac{1}{2}$ it is anti-persistent. The argument is simple \cite{Man}. Let us compute the auto-correlation function 
\begin{align}
C(\tau)&= \frac{\left \langle (Q(t)-Q(t-\tau))(Q(t+\tau)-Q(t)) \right\rangle}{\left \langle (Q(t+\tau)-Q(t))^2 \right\rangle} \cr 
         &= 2^{2\mathcal{H}-1} - 1 \ ,
\end{align}
where we have used \eqref{Hurstd} with $q=2$, i.e. $\langle (Q(t+\tau)-Q(t))^2 \rangle = C_2\tau^{2\mathcal{H}}$. Since $C(\tau)$ governs the persistence of memory, it leads to the above interpretation.
The bound $\mathcal{H}<1$, or $\mathcal{Z}>1$, has been observed in the gravity before (c.f. the discussion in section 3.1 and its relation to the unitarity bound.)
The Reggeon field theory in $\epsilon$ expansion predicts $\mathcal{H} = 0.57$ for $d=2$ and $\mathcal{H}=0.64$ for $d=1$, which is consistent with the observation that the market is usually persistent in memory.

The gravitational dual theory by itself cannot tell which exponent should be realized in nature or in market. The analogue of the holographic running might be important from the ``UV" critical exponent $\mathcal{Z} =2$ ($\mathcal{H}=\frac{1}{2}$) to that for the ``IR" exponent. By introducing two types of massive vector fields with different mass, which is related to $\mathcal{Z}$, we may obtain the geometry interpolating different dynamical critical exponents. 
 On the other hand, the experimental determination of the Hurst exponent from the market data is a challenging subject. See \cite{nr}\cite{Man} for some related discussions. We just mention that because the non-trivial Hurst exponent typically indicates the violation of the Martingale or the efficient market hypothesis,\footnote{This is not necessarily the case especially when the distribution does not have stationary increment. See e.g. \cite{he}. A similar argument could be stated in field theories.  Non-local but free field theories such as $S = \int dw d^dk\phi^\dagger (w^{2\mathcal{H}}-k^2)^\nu \phi$ known as generalized free field theories give non-trivial dynamical critical exponent, but they are rather trivial. In such theories, higher cumulant does not contain any non-trivial information at all.} we expect the Hurst exponent or non-linear fixed point is rather unstable and cannot last forever.

\section{Discussions}
In this paper, we proposed novel applications of non-relativistic gauge/gravity correspondence. One is the Reggeon field theory and another is the non-linear quantum finance. The relation between the two has been suggested some time ago \cite{quant2}, and we have made more concrete proposals in this paper. We have seen that the strongly coupled regime of the both theories is beyond the scope of the perturbative field theories, and the dual gravity computation is promising. We have proposed the two-point functions from the gravity computation.


The quantum finance is base on the relation between stochastic systems and quantum field theories. We believe that the stochastic process in gauge/gravity correspondence is worth studying further. Recently, an effective Langevin equation for heavy charged particles in the quark-gluon plasma has been computed by using the AdS/CFT correspondence  \cite{deBoer:2008gu}\cite{Son:2009vu}\cite{Giecold:2009cg}, where they have pointed out its relation to black hole Hawking radiation. Introducing external potential in their setup, we can study the geometric Brownian motion of stochastic strings, which may be relevant in quantum finance. The random noise predicted in AdS/CFT is not white, so when it turns out to be solvable (possibly by studying the dual gravity regime), it might become a novel non-Gaussian market model. 

On the theoretical side, it would be interesting to find a way to convert their effective Langevin equations into the form of non-relativistic field theories as we have presented in section 3.1. In this way, we may be able to find a clue to derive our proposed phenomenological gravity background from the string theory. All in all, there is no concrete examples of non-relativistic gauge/gravity correspondence whose gauge side is identified with the gravity side (and vice versa), so this approach could be a major breakthrough if successful \cite{wip}.

We would like to conclude this paper with a few words about our philosophy of applying gauge/gravity correspondence to non-linear finance. Much like in the cold atoms or unitary fermion systems, there is currently no theoretical foundation that the gravity description is suitable for describing the financial market except for general symmetry arguments and its validity in the free theory limit. From the phenomenological model building perspective, there is nothing wrong about proposing a ``solvable" model from completely different perspective. Indeed, it might be better suited here than the condensed matter application because we do not know the fundamental principle of market unlike in the condensed matter systems. 
On the other hand, the Reggeon field theory {\it does} have its origin in QCD, so it would be wonderful to derive the phenomenological non-relativistic background discussed in this paper from the AdS/QCD correspondence whose theoretical foundation is much firmer.

\section*{Acknowledgements}
The author would like to thank Soo-Jong Rey for suggesting AdS/Finance correspondence. The work was supported in part by the National Science Foundation under Grant No.\ PHY05-55662 and the UC Berkeley Center for Theoretical Physics.

\appendix

\section{Scale invariance vs conformal invariance in Galilean invariant theories}
In this appendix, we discuss the energy-momentum tensor of Galilean invariant field theories to study the condition when the scale invariance indicates the conformal invariance. The non-relativistic energy-momentum tensor with the conformal invariance has been studied in \cite{Jackiw:1990mb}.
We assume the translational invariance in time and space, which means we have a conserved energy-momentum tensor
\begin{align}
\partial_t T^{0i} + \partial_j T^{ji} &= 0 \cr
\partial_t T^{00} + \partial_i T^{i0} &= 0 \ 
\end{align}
corresponding to 
\begin{eqnarray}
H = \int d^dx T^{00} \ , \ \ P^i = \int d^dx T^{0i} \ .
\end{eqnarray}
The spatial rotational invariance demands that the energy-momentum tensor be symmetric $T^{ij} = T^{ji}$ (note, however, $T^{0i} \neq T^{i0}$ in general).

We also assume that the theory is invariant under the Galilean boost by demanding that the $U(1)$ particle number density is related to the energy-momentum tensor
\begin{eqnarray}
m\dot{\rho} = -\partial_i T^{0i} \ . \label{partn}
\end{eqnarray}
Then the Galilean boost density $\mathcal{G}^i = t T^{0i} -mx^i\rho$ satisfies
\begin{eqnarray}
 \partial_t \mathcal{G}^i + \partial_j(tT^{ij} - x^iT^{0j}) = 0 \ . \label{gal}
\end{eqnarray}
The corresponding conserved charge is $G^i = \int d^d x \mathcal{G}^i$

Now suppose that the energy-momentum tensor satisfies the following condition
\begin{eqnarray}
2T^{00} - T^{ij}\delta_{ij} = \partial_t \mathcal{S} + \partial_j \mathcal{A}^j \ , \label{dila}
\end{eqnarray}
then one can show that the dilatation density
\begin{eqnarray}
\mathcal{D} = tT^{00} -\frac{1}{2}x_iT^{0i} - \frac{\mathcal{S}}{2} 
\end{eqnarray}
is conserved. The corresponding charge is given by $ D = \int d^d x \mathcal{D}$. Thus, the condition \eqref{dila} is the requirement of scale invariance. Note that one can always redefine $2T^{00} \to 2T^{00} + \partial_j \mathcal{A}^j$ to remove $\mathcal{A}^j$, so only the non-trivial condition is the existence of $\mathcal{S}$. Furthermore, if $\mathcal{S}$ is a total divergence such that $\mathcal{S} = \partial_i \sigma^i$, then we can improve the energy-momentum tensor as $T^{00} \to T^{00} + \partial_j\partial_t \sigma^j + \partial_i \mathcal{A}^i$ so that the right hand side of \eqref{dila} is zero.

The condition that the right hand side of \eqref{dila} can be improved to be zero is an analogue of the traceless condition for the energy-momentum tensor in relativistic field theories \cite{Polchinski:1987dy}. As in the relativistic case, \eqref{dila} indicates, if the right hand side vanishes, an additional conserved density
\begin{eqnarray}
\mathcal{K} = t^2 T^{00} - t x_i T^{0i} + \frac{m}{2}x^2\rho \ , \label{confg}
\end{eqnarray}
which generates non-relativistic conformal transformation whose charge is $K = \int d^d x \mathcal{K}$.
 In order to show that \eqref{confg} is conserved, the Galilean invariance \eqref{partn} and the trace condition on the energy-momentum tensor are crucial. 

The discussion here states that the scale invariance and Galilean invariance do not necessarily imply the conformal invariance in non-relativistic field theories, much like in relativistic field theories. The criterion of the conformal invariance is whether we could improve the energy-momentum tensor so that the trace condition $2T^{00} - T^{ij}\delta_{ij} = 0$ is satisfied. A non-trivial possibility of $\mathcal{S}$ which is not a divergence of another current is the obstruction. As far as we know, however,  there is no known physical example of non-trivial $\mathcal{S}$ in literatures.

Let us do a no-relativistic version of the exercise done in \cite{Polchinski:1987dy}. Consider the non-relativistic action
\begin{eqnarray}
S = \int dt d^d x \left(i\phi_a^\dagger \partial_t \phi_a - \frac{1}{2m} \partial_i\phi^\dagger_a \partial_i \phi_a - \frac{\lambda_{abcd}}{4} \phi_a\phi_b\phi^\dagger_c\phi^\dagger_d\right) \ 
\end{eqnarray}
in $d= 2-\epsilon$ dimension. The reality of the action demands $\lambda_{abcd} = \lambda^*_{cdab}$.
To the first order in $\epsilon$, the candidate for $\mathcal{S}$ and $\mathcal{A}^j$ is given by
\begin{align}
\mathcal{S} &= i\xi^{ab} \phi_a^\dagger\phi_b \cr 
\mathcal{A}^j &= \frac{\xi^{ab}}{2m}\left(\phi_a^\dagger\partial^j \phi_b + \phi_a\partial^j\phi_b^\dagger \right)\ ,
\end{align}
with (real) antisymmetric $\xi^{ab}$. At one-loop, the scale invariance demands
\begin{align}
-\epsilon \lambda_{abcd} + \frac{1}{16\pi^2}\lambda_{abef}\lambda_{efcd} + \xi^{mc}\lambda_{abmd} + \xi^{md}\lambda_{abmc} - \xi^{am}\lambda_{mbcd} -\xi^{bm}\lambda_{macd} = 0 \ \label{si}
\end{align}
while the conformal invariance demands
\begin{align}
 -\epsilon \lambda_{abcd} + \frac{1}{16\pi^2}\lambda_{abef}\lambda_{efcd}  = 0  \ . \label{cf}
\end{align}
The latter is in principle stronger than the former because of the additional unknowns $\xi^{ab}$. We can show, however, that \eqref{si} implies \eqref{cf} by contracting \eqref{si} with $\xi^{ma}\lambda_{cdmb} + \xi^{mb}\lambda_{cdma} - \xi^{cm}\lambda_{mdab} -\xi^{dm}\lambda_{mcab}$. Thus, the scale invariance and Galilean invariance does suggest the conformal invariance in $\lambda |\phi|^4$ non-linear Schr\"odinger theory.

Let us finish this appendix with another peculiar example in $(1+0)$ dimension. Consider the non-relativistic Liouville-like Lagrangian
\begin{align}
 L &= i \phi^\dagger \partial_t \phi - \mu e^{|\phi|^2} \cr
   &= -\rho\partial_t\theta -\mu e^{\rho}
\end{align}
where we have introduced the polar coordinate $\phi = \sqrt{\rho} e^{i\theta}$.  The theory is scale invariant under $\rho \to \rho + \lambda$.  Note that if it were defined in $(1+d)$ dimension ($d>0$) with additional kinetic terms, the theory would be Galilean invariant (but not scale invariant). The Hamiltonian $T^{00} = \mu e^{\rho} = -\partial_t \theta$ is a time derivative, so the dilatation charge $\mathcal{D} = t\mu e^{\rho} + \theta$ is conserved. However, since $\theta$ is not a derivative of something else, we cannot construct a conserved non-relativistic conformal charge. This example might suggest a possibility to construct a counterexample, but we have not come up with any in higher dimensions with Galilean invariance.

\section{Non-relativistic conformal algebra}
We summarize the non-relativistic conformal algebra for $d=2$. For $d=1$, there is no angular momentum $J$, and no indices for $P$ and $G$ in the following commutation relations.
\begin{align}
i[J,P] &= -i P \ , \ \ i[J,\bar{P}] = i \bar{P} \ , \ \ i[J,G] = - iG \ , \ \ i[J,\bar{G}] = i\bar{G} \ ,\cr
i[H,G] &= P \ , \ \ i[H,\bar{G}] = \bar{P} \ , \ \ i[K,P] = - G \ ,  \ \ i[K,\bar{P}] = - \bar{G} \ ,\cr
i[D,P] &= - P \ , \ \ i[D,\bar{P}] = - \bar{P} \ , \ \ i[D,G] = G \ , \ \ i[D,\bar{G}] = \bar{G} \ ,\cr
 i[D,H] &= -2H \ , \ \ i[H,K] = D \ , \ \ i[D,K] = 2K \ , \ \ i[P,\bar{G}] = 2M \ , 
\end{align}
where $J = J_{12}$ is $U(1)$ angular momentum, $P = P_1+iP_2$ and $\bar{P} = P_1 - iP_2$ are spatial momenta, $H$ is the Hamiltonian, $G = G_1+iG_2$ and $\bar{G} = G_1-iG_2$ are Galilean boost, $D$ is the dilatation, $K$ is the special conformal transformation, and $M$ is the number density operator which is the center of the non-relativistic conformal algebra. 

The representation theory of the non-relativistic conformal algebra relevant for the non-relativistic conformal field theories can be found in \cite{Dobrev}\cite{Nakayama:2008qm}.

\section{Confluent Hypergeometric function}
The confluent Hypergeometric function $U(a,b;x)$ is a solution of Kummer's equation
\begin{eqnarray}
x\frac{d^2 U}{dx^2} + (b-x) \frac{dU}{dx} - a x = 0 \ 
\end{eqnarray}
with the series expansion
\begin{align}
U(a,b;x) = &x^{1-b} \left[\frac{\Gamma(-1+b)}{\Gamma(a)} + \frac{(-1-a+b)\Gamma(-2+b)}{\Gamma(a)} x + \cdots  \right] \cr
 &+ \frac{\Gamma(1-b)}{\Gamma(1+a-b)} - \frac{a\Gamma(-b)}{\Gamma(1+a-b)} x + \cdots \ .
\end{align}

Alternatively it has an integral representation
\begin{eqnarray}
U(a,b;x) = \frac{1}{\Gamma(a)} \int_0^\infty dt e^{-xt} t^{a-1}(1+t)^{b-a-1}  \ .
\end{eqnarray}
It satisfies 
\begin{eqnarray}
U(a,b;x) = z^{1-b}U(1+a-b,2-b;x) \ .
\end{eqnarray}

\end{document}